\documentclass[useAMS,usenatbib]{mn2e}
\usepackage{epsfig,natbib}
\usepackage{amssymb}
\usepackage{graphicx}
\usepackage{placeins}
\usepackage{times}
\usepackage{epsfig}

\newcommand\ion[2]{#1{\sc #2}}%  ion, i.e. CII = \ion{C}{ii}
\title[The Galaxy Counterparts of the two high-metallicity DLAs towards Q\,0918+1636] 
{On the two high-metallicity DLAs at $z=2.412$ and $z=2.583$ towards Q\,0918+1636
\thanks{Based on observations carried out under prog.  ID 084.A-0303 and
089.A-0087 with the X-Shooter spectrograph installed at the Cassegrain focus of
the Very Large Telescope (VLT), Unit 2 -- Kueyen, operated by the European
Southern Observatory (ESO) on Cerro Paranal, Chile. Based on observations made
with the NASA/ESA Hubble Space Telescope, obtained at
the Space Telescope Science Institute, which is operated by the Association of
Universities for Research in Astronomy, Inc., under NASA contract NAS 5-26555.
These observations are associated with program 12553. Based on observations made with the Nordic Optical Telescope, operated
on the island of La Palma jointly by Denmark, Finland, Iceland,
Norway, and Sweden, in the Spanish Observatorio del Roque de los
Muchachos of the Instituto de Astrofisica de Canarias.}}
\author[Fynbo et al.]{
J. P. U. Fynbo$^{1}$\thanks{E-mail:jfynbo@dark-cosmology.dk},
S. J. Geier$^{1,2}$,
L. Christensen$^{1}$,
A. Gallazzi$^{3,1}$,
J.-K. Krogager$^{1,4}$,\newauthor
T. Kr\"uhler$^{1}$,
C. Ledoux$^{4}$,
J. R. Maund$^{5,6}$,
P. M\o ller$^{7}$,
P. Noterdaeme$^{8}$,\newauthor
T. Rivera-Thorsen$^{1,9}$,
M. Vestergaard$^{1,10}$
\\
$^{1}$Dark Cosmology Centre, Niels Bohr Institute, Copenhagen University, Juliane Maries Vej 30, 2100 Copenhagen O, Denmark\\
$^{2}$Nordic Optical Telescope, Apartado 474, 38700~Santa Cruz de La Palma, Spain\\
$^{3}$INAF -- Osservatorio Astrofisico di Arcetri, Largo Enrico Fermi 5, 50125 Firenze, Italy,\\
$^{4}$European Southern Observatory, Alonso de C\'ordova 3107, Vitacura, Casilla 19001, Santiago 19, Chile\\
$^{5}$Astrophysics Research Centre, School of Mathematics and Physics, Queen's University Belfast, Belfast BT7 1NN, UK\\
$^{6}$Royal Society Research Fellow,\\
$^{7}$European Southern Observatory, Karl-Schwarzschildstrasse 2, 85748 Garching bei M\"unchen, Germany\\
$^{8}$Institut d'Astrophysique de Paris, CNRS-UPMC, UMR7095, 98bis bd Arago, 75014 Paris, France\\
$^{9}$Department of Astronomy, AlbaNova, Stockholm University, 106 91 Stockholm, Sweden\\
$^{10}$Steward Observatory, University of Arizona, 933 North Cherry Avenue, Tucson, AZ 85721\\
}

\begin{document}

%\date{Accepted . Received ; in original form}

%\pagerange{\pageref{firstpage}--\pageref{lastpage}} \pubyear{2010}
\pagerange{1 -- 1} \pubyear{2013}

\maketitle

%\label{firstpage}

\begin{abstract}
The quasar Q0918+1636 ($z=3.07$) has two intervening high-metallicity Damped
Lyman-$\alpha$ Absorbers (DLAs) along the line of sight, at redshifts of $z=2.412$
and 2.583. The $z=2.583$ DLA is located at a large impact parameter of 16.2 kpc,
and despite this large impact parameter it has a very high metallicity
(consistent with solar), a substantial fraction of H$_2$ molecules, and it is
dusty as inferred from the reddened spectrum of the background QSO. The $z=2.412$
DLA has a metallicity of $[M/H]=-0.6$ (based on \ion{Zn}{ii} and \ion{Si}{ii}). 
In this paper we present new observations of this interesting sightline
consisting of deep multi-band imaging and further VLT spectroscopy. By fitting
stellar population synthesis models to the photometric SED we constrain the
physical properties of the $z=2.583$ DLA galaxy, and we infer its morphology by
fitting a Sersic model to its surface brightness profile. We find it to be a
relatively massive ($M_{\star} \approx 10^{10}~M_{\odot}$), strongly
star-forming (SFR $\approx 30~\mathrm{M}_{\odot}$~yr$^{-1}$), dusty
($E_{B-V}=0.4)$ galaxy with a disk-like morphology.  We detect strong emission
lines from the $z=2.583$ DLA ([\ion{O}{ii}]\,$\lambda$3727,
[\ion{O}{iii}]\,$\lambda$4960, [\ion{O}{iii}]\,$\lambda$5007, H$\beta$, and
H$\alpha$, albeit at low signal-to-noise (S/N) ratio except for the
[\ion{O}{iii}]\,$\lambda$5007 line). The metallicity derived from the emission
lines is consistent with the absorption metallicity ($12 + \log(\rm{O/H}) = 
8.8\pm0.2$). We also detect
[\ion{O}{iii}]\,$\lambda$5007 emission from the galaxy counterpart of the
$z=2.412$ DLA at a small impact parameter ($<2$ kpc).  Overall our findings are
consistent with the emerging picture that high-metallicity DLAs are associated
with relatively luminous and massive galaxy counterparts, compared to typical
DLAs.
\end{abstract}

\begin{keywords}
   galaxies: formation
-- galaxies: high-redshift
-- galaxies: ISM
-- quasars: absorption lines
-- quasars: individual: SDSS J\,091826.16$+$163609.0
-- cosmology: observations
\end{keywords}

\section{Introduction}

%The study of galaxies in absorption against the light of background QSOs was
%for a long time the only available for redshifts larger than about unity
%\citep[e.g.,][]{1981ARA&A..19...41W,2005ARA&A..43..861W}. Then, from the 2nd
%half of the 1990ies, the study of high-$z$ galaxies in emission went through a
%break-through that is still unfolding
%\citep[e.g.,][]{2002ARA&A..40..579G,2011ARA&A..49..525S}. However, bringing
%together the information from absorption and emission studies is still a
%relatively poorly developed field. Despite some progress
%\citep[e.g.,][]{2002ApJ...574...51M} we still have less than a dozen examples
%galaxy counterpars of absorption selected galaxies \citep{2012MNRAS.424L...1K}.

For a long time the only available method for studying galaxies at redshifts
$z>1$ (barring QSO host galaxies) was to look at them in absorption against the light of
background QSOs \citep[e.g.,][]{1981ARA&A..19...41W,2005ARA&A..43..861W}.
Then, from the second half of the 1990ies, the study of high-$z$ galaxies in
emission went through a breakthrough that is still unfolding
\citep[e.g.,][]{2002ARA&A..40..579G, 2011ARA&A..49..525S}. However, combining
the information from absorption and emission lines is still a poorly developed
field. Although more than 10\,000 of the so-called Damped Lyman-$\alpha$
Absorbers (DLAs) have been found so far \citep{2012A&A...547L...1N}, and
despite some progress \citep[e.g.,][]{2002ApJ...574...51M} in finding their
galaxy counterparts, we still have less than a dozen examples of such
absorption selected galaxies \citep[][see also Rauch et al.\ 2008,
Rauch \& Haehnelt 2011 and Schulze et al.\ 2012]{2012MNRAS.424L...1K}.

Expanding the sample is of great interest as we in this manner obtain unique
information about the kinematics and chemical composition of gas surrounding
the central $\sim$1 kpc which are typically studied in emission. This information is
vital for probing current ideas about the role of processes like 
inflow of pristine gas and outflow of enriched gas in galaxy formation 
and evolution \citep[e.g.,][and references therein]{2009Natur.457..451D,
2010ApJ...718.1001B,2011MNRAS.418.1796F,2013MNRAS.432..455D,Crighton13}.

The $z=3.07$ quasar SDSSJ\,091826.16$+$163609.0 was selected in the survey for
high-metallicity DLAs described in \citet{2010MNRAS.408.2128F} and
\citet{2011MNRAS.413.2481F}.  It was selected due to the presence of a DLA at
$z=2.412$ with strong \ion{Fe}{ii} lines.  After obtaining deep X-Shooter
spectroscopy of the QSO, \citet{2011MNRAS.413.2481F} serendipitously discovered
a second DLA at $z=2.583$ along the line of sight with even stronger metal
lines.  \citet{2011MNRAS.413.2481F} detected the forbidden [\ion{O}{ii}] and
[\ion{O}{iii}] emission lines of the galaxy counterpart of this second DLA.
The galaxy is located at an impact parameter of $2\arcsec$, corresponding to
$\sim16$ kpc at $z=2.583$.  The $z=2.412$ DLA was not detected in emission in
that study.

In this paper, we present new results based on new observations of this
sightline obtained with the Hubble Space Telescope ({\it HST}), ESO Very Large
Telescope (VLT) and Nordic Optical Telescope (NOT).  In
Sect.~\ref{data} we give an overview of the observations and data reduction,
and describe the data analysis and the results in Sect.~\ref{z2.58} and
Sect.~\ref{z2412}.  Finally, Sect.~\ref{Discussion_Section} contains a
discussion of our
findings and their implications for the field. % shootout ;)

Throughout this paper, we use a flat $\Lambda$ CDM cosmology with
$\Omega_{\Lambda}=0.728$, $\Omega_m=0.272$ and a Hubble constant of
$H_0=70.4$~km~s$^{-1}$~Mpc$^{-1}$ \citep{2011ApJS..192...18K}.
All magnitudes are given in the AB system.

\section{Observations and data reduction}
\label{data}

\subsection{HST imaging}
\label{hst}

The field of Q\,0918+1636 was observed with the Wide Field Camera 3 (WFC3) on the HST
on two epochs in November 2011 (with the NIR detector in the F105W and F160W
filters) and on April 18 2012 (with the UVIS detector in the F606W filter). The
roll-angle of the telescope was set such that the $z=2.583$ DLA galaxy falls
between the diffraction spikes of the Point Spread Function (PSF) of the QSO.
The two observations with the NIR detector were taken using the
WFC3-IR-DITHER-BOX-MIN pattern providing an optimal 4-point sampling of the
PSF. The UVIS observation was taken using the WFC3-UVIS-DITHER-BOX pattern.

We have reduced and combined the images using the software package {\tt
multidrizzle} provided by the STScI. By shifting and combining the images taken
with sub-pixel offsets one achieves a better sampling of the PSF, which in the
case of the IR observations is crucial as the PSF is poorly sampled in
the native 0\farcs13~px$^{-1}$ images. %Furthermore, the drizzle-algorithm
%allows us to reduce the pixel size in the combination of the images providing
%sharper and more well-sampled images. 
For this work we have set the parameter
{\tt pixfrac} to $0.7$ in all reductions and used a final pixel scale of
0\farcs06~px$^{-1}$ for IR and 0\farcs024~px$^{-1}$ for UVIS.  For a detailed
description of the parameters in the software we refer to the {\tt multidrizzle} user manual.

\begin{table}
\caption{Log of observations\label{table:log}}
\begin{center}
\begin{tabular}{llr}
\hline
\hline
Band  & Obs. Date & Exp. Time \\
      &            &   (sec)   \\
\hline
{\it HST}/WFC3/F105W & Nov 9 2011 & 2612 \\ 
{\it HST}/WFC3/F160W & Nov 9 2011 & 2612 \\ 
{\it HST}/WFC3/F606W & Apr 9 2011 & 2523 \\ 
\hline
NOT/Alfosc/$g$ & Jan 25--26 2012 & 8400 \\
NOT/Alfosc/$u$ & Jan 25--26 2012 & 11500\\
NOT/NOTCam/$Ks$ & Mar 3 2012 & 7830 \\
\hline
VLT/X-Shooter$^{1}$ stare PA=0$^\mathrm{o}$ & Feb 16 2010 & 3600 \\
VLT/X-Shooter$^{1}$ stare PA=60$^\mathrm{o}$ & Feb 16 2010 & 3600 \\
VLT/X-Shooter$^{1}$ stare PA=$-60^\mathrm{o}$ & Feb 16 2010 & 3600 \\
VLT/X-Shooter nod PA=$-66^\mathrm{o}$& Apr 15 2012 & 2920 \\
VLT/X-Shooter stare PA=162$^\mathrm{o}$& Apr 15 2012 & 6400 \\
VLT/X-Shooter nod PA=66$^\mathrm{o}$ & Mar 15--16 2013 & 10800 \\
\hline
\hline
\end{tabular}
\end{center}
$^{(1)}$ Already published in \citet{2011MNRAS.413.2481F}.
\end{table}

%We obtain the photometric zeropoints in the AB magnitude system from
%the FITS header keywords PHOTPLAM and PHOTFLAM using the prescription
%in the WFC3 handbook.

%$ZP=-2.5\cdot(\log(photplam^{2}\cdot~photflam)-\log(3\cdot~e^{18})\cdot\ln10^{-1})-48.6$

\subsection{NOT imaging}
\label{nodata}

On the nights of Jan 25--26 2012 and March 3 2012 Q\,0918$+$1636 was observed
with the Andalucia Faint Object Spectrograph and Camera (ALFOSC) and with the
Nordic Optical Telescope near-infrared Camera and spectrograph (NOTCam) at the
NOT. A total of 8400~s, 11500~s and 7830~s of exposure time was obtained in the
$g$-band, $u$-band, and
% 1080s in the $Y$-band, 2700s in the $J$- and $H$-band each,
$K_\mathrm{s}$-band, respectively (see Table~\ref{table:log}). Observing
conditions were clear, with an average seeing FWHM of $\sim1\arcsec$ in the
January 2012 % thin clouds and seeing $\sim2\arcsec$ in the February
nights, and clear and sub-arcsec seeing on March 3rd 2012.% all three nights.
The optical images were reduced using IRAF\footnote{IRAF is distributed by the
National Optical Astronomy Observatory, which is operated by the Association of
Universities for Research in Astronomy (AURA) under cooperative agreement with
the National Science Foundation.} standard procedures. The NOTCam images were
reduced with custom IDL scripts, using a running-median for sky-subtraction,
and an object mask for 2nd pass sky-subtraction. The NOTCam distortion
correction was applied using the {\sc iraf}/{\tt geotran} task.

\subsection{VLT/X-Shooter spectroscopy}
\label{shoot}

%Further
Q\,0918$+$1636 was observed with VLT/X-Shooter on April 15
2012. Both a stare observation at a position angle of 162$^\mathrm{o}$ East of
North and a nodding observation at a position angle of $-66$$^\mathrm{o}$ East
of North were obtained. The observation at position angle $-66$$^\mathrm{o}$
East of North was a mistake: it should have been at $66$$^\mathrm{o}$ East of
North with the purpose of covering the $z=2.583$ DLA galaxy and the QSO.  In a
stare observation the target is kept at a fixed position on the slit throughout
the observation, whereas in a nodding observation an observing block
consists of four exposures between which the target is moved in an ABBA pattern
along the slit. A deeper observation was obtained on March 15--16 2013 at the
correct position angle of 66$^\mathrm{o}$ East of North for covering the
$z=2.583$ DLA galaxy and the QSO using the nodding template with a nod-throw of
$4\arcsec$. In Fig.~\ref{fig:slits} we show the orientation of the slits in all the X-Shooter observations,
both the new observations reported here and the previous observations of
\citet{2011MNRAS.413.2481F}.  The purpose of
the stare observation at position angle of 162$^\mathrm{o}$ East of North was
to determine the redshift of the galaxy seen in the bottom of
Fig.~\ref{fig:slits} in order to establish if this could be the galaxy
counterpart of the $z=2.412$ DLA.  For the full log of observations we refer to
Table~\ref{table:log}.

The spectra from March 2013 were reduced with the ESO X-Shooter pipeline 2.0
\citep{2010SPIE.7737E..56M,2011AN....332..227G}.  We use the default parameters
for the first five recipes which perform the basic calibrations (master darks, order
prediction, flat fields, and the 2D maps for later rectification of the
spectra).  For the reduction of the object frames we use the corresponding
pipeline recipes for the stare and nodding modes, with parameters optimized to
provide the best possible sky-subtraction. The flux standard star LTT3218 was
observed in both nights. Those spectra were reduced with the same calibration
data as the spectra of the QSO/DLA system, and sampled onto the same spatial
and wavelength grid. The extracted 1-dimensional (1D) standard star spectra were 
divided by the known tabulated spectrum of the standard star (which is first interpolated to the
same wavelength grid), and the result smoothed with a kernel of 30 pixels to
obtain a clean response curve. Each individual 2D spectrum was first
normalized to an integration time of 1s, and then divided by the also
normalized response curve from the corresponding night. The resulting
flux-calibrated spectra from the 3 individual observing blocks (each comprising
one hour of integration time) were collapsed along the wavelength axis to
determine the peak positions of their spectral PSFs (SPSFs), and subsequently aligned
on the spatial axis. Thereafter, they were stacked by means of a median
combination.  Galactic extinction corrections were taken from
\citet{2011ApJ...737..103S}\footnote{provided by {\tt http://ned.ipac.caltech.edu/}}
and implemented with the $fm\_unred$ code in IDL.
The final SPSF was determined by collapsing the error-weighted 2D stack along the wavelength axis in the $H$-band wavelength range and the 1D spectrum of the
quasar itself was extracted by applying the corresponding normalized weights,
similar to the optimal extraction procedure described in
\cite{1986PASP...98..609H}. %The extraction window spans a spatial extension of
%$1\farcs8$ around the  centre of the trace.  A 1d error spectrum is extracted
%the following way: with NSP being the normalized SPSF along the extraction
%window, the 1d variance is calculated as \begin{equation}
%Var(\lambda)=\frac{\sum_{y=y1}^{y2}Err(\lambda,y)^{2}*NSP(y)^{2}}{(\sum_{y=y1}^{y2}
%NSP(y)^2)^2} \end{equation} and the 1d errors are then the square root of that
%variance.  
To check the flux level we integrated the 1D spectrum over the transmission
curves of the NOTCam $J$- and $H$-band filters, and compared with the
photometry from \cite{2013ApJS..204....6F}. The measurements agree within the
errorbars, thus no correction of the flux level was necessary.  With a gaussian
fit to the SPSF we determined the seeing of the combined spectrum to be
$\sim0\farcs8$ in the $H$-band. 

%\section{Results}
%\label{results}

\section{The $\lowercase{z}=2.583$ DLA galaxy}
\label{z2.58}

%In this section we present our analysis of the $z=2.583$ DLA and its galaxy counterpart.

\subsection{HST imaging}
\label{out_of_hst}

Based on the high resolution HST imaging we can now improve the relative
astrometry over that presented in \citet{2011MNRAS.413.2481F}. We find that
the DLA galaxy is located at an impact parameter of $1.98\pm0.02$ arcsec
from the QSO at a position angle of $-115^\mathrm{o}$ East of North,
consistent with the earlier measurements. This impact parameter corresponds to a proper distance of 16.2 kpc at
$z = 2.583.$

\begin{figure}
\includegraphics[width=0.48\textwidth]{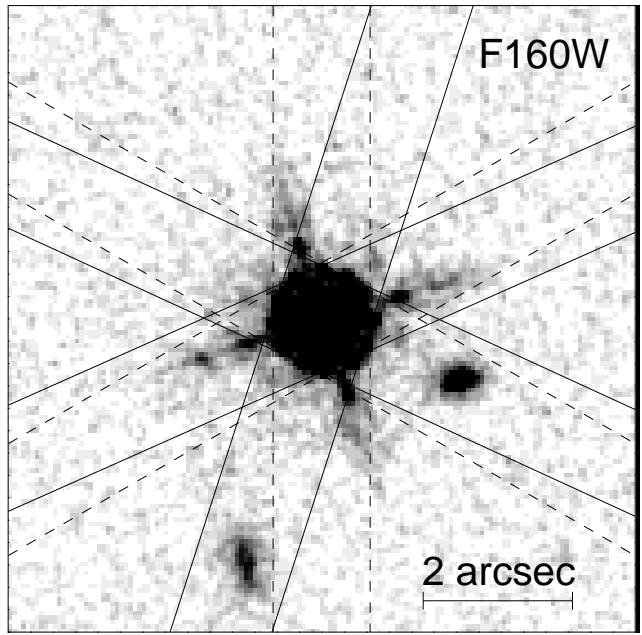}
\caption{The {\it HST}/WFC3 F160W image and with all the X-Shooter slit
orientations indicated with dashed lines for the observations presented in
\citet{2011MNRAS.413.2481F} and full-drawn lines for the new observations
presented here (see also Table~\ref{table:log}).
\label{fig:slits}}
\end{figure}

We used the G{\small ALFIT} tool \citep{2002AJ....124..266P,2010AAS...21522909P} to fit
2D Sersic models, convolved with the PSF, to the HST images of the DLA galaxy.
The Point Spread Functions (PSFs) for the {\it HST} images were simulated using
the software {\sc TinyTim}. We chose to simulate the PSFs instead of using an
empirical PSF as the model PSF has higher S/N ratio in the outer
parts, where the PSFs from the data have high noise due to the background.  We
did the PSF simulation by first creating models using {\sc TinyTim} for each
position of the target in the four-point dither pattern, assuming a QSO
spectrum for the wavelength dependent PSF modelling, and taking into account
the aberrations of the telescope as specified in the auxiliary data files. The
models were sub-sampled by a factor of 5 compared to the native pixel scale of
the detectors in order to position the model PSF more accurately. We then
re-sampled the model PSF images to the native sampling and convolved them with
the appropriate filter-specific Charge Diffusion Kernel. The four "raw" PSF
images were then combined by the pipeline task {\tt multidrizzle} in {\sc iraf}
using the same parameters as for the data reduction. This allows us to mimic
the effects of the reduction procedures. The results from G{\small ALFIT} are the best-fit values for the effective
half-light radius $r_{\rm eff}$, the Sersic index $n$, and the axis ratio
$\frac{b}{a}$), which quantify the structure of the galaxy. 
The circularized radius is calculated as $r_{c}=r_{\rm eff}\cdot\sqrt{\frac{b}{a}}$.

G{\small ALFIT} also delivered photometry in all three bands, summarized in Table
\ref{Galfit_results}. Galactic extinction corrections are taken from the
\cite{2011ApJ...737..103S} maps.

\begin{figure*}
\includegraphics[width=0.95\textwidth]{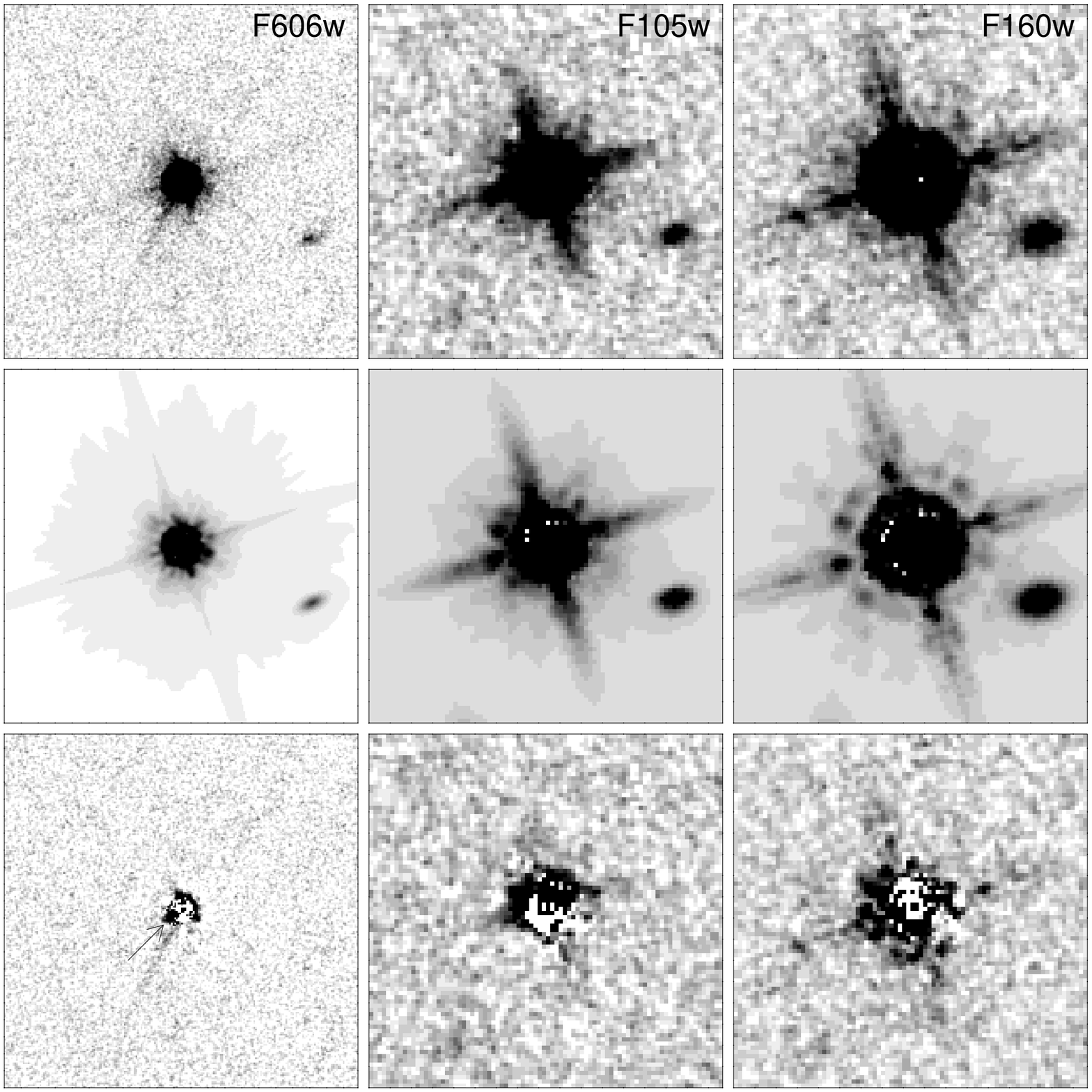}
\includegraphics[width=0.95\textwidth]{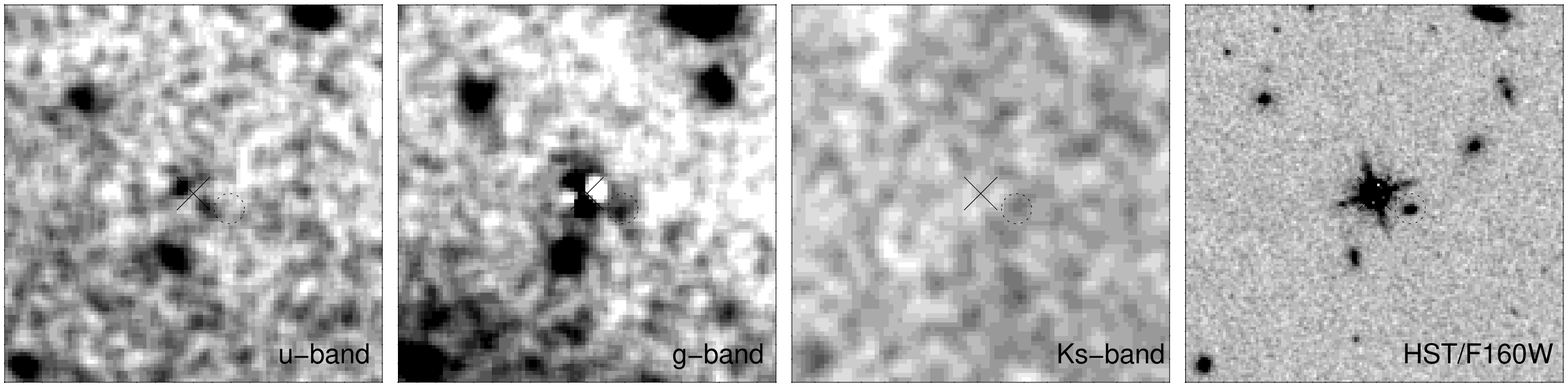}
\caption{Top 9 panels: 5$\times$5 arcsec$^2$ sections centred on the QSO in
F606W, F105W and F160W (from left to right).  The DLA galaxy is seen south-west of
the QSO at an impact parameter of $2\arcsec$. The top row shows the science
images, the middle row shows the G{\small ALFIT} model and the bottom row shows the
residuals after subtracting the model (QSO PSF and galaxy models) from the
science images. The field size is 5$\times$5 arcsec$^2$ and the images are
oriented with North up and East to the left. The arrow in the lower left panel
indicates the possible signature of the $z=2.412$ DLA galaxy. The tail 
extending below the QSO is a result of Charge Transfer Inefficiency.
Bottom panel: 19$\times$19 arcsec$^2$ sections centred on the QSO in the u-, g-, and 
Ks-bands after PSF-subtraction.
Also shown to the right is the {\it HST}/F160W 
band image on the same scale for comparison. The position of the $z=2.583$
DLA galaxy is marked with a dotted circle. Again the images are oriented with
North up and East to the left.
\label{fig:mosaic}}
\end{figure*}

\begin{table*}
\caption{Results from the G{\small ALFIT} fits and NOT photometry. $b/a$
is the ratio of the minor and major axis radii and $n$ is the 
Sersic index.
The magnitudes in the HST filters were computed by G{\small ALFIT}, whereas aperture photometry was done on the NOT 
images.\label{Galfit_results}}
\begin{center}
\begin{tabular}{lllllll}
\hline
\hline
Band  & mag &  r$_{\rm eff}$ & PA & b/a & n & r$_{c}$ \\
      & (AB) & (kpc) & (deg) & & & (kpc) \\
F606W & 25.46$\pm$0.13 & 1.2$\pm$0.2 & $-37\pm$6 & 0.4$\pm$0.1 & 0.8$\pm$0.4 & 0.73 \\
F105W & 24.61$\pm$0.09 & 1.3$\pm$0.3 & $-44\pm$9 & 0.3$\pm$0.2 & 0.1$\pm$0.7 & 0.69 \\
F160W & 23.68$\pm$0.06 & 1.4$\pm$0.1 & $-47\pm$6 & 0.4$\pm$0.1 & 1.1$\pm$0.4 & 0.93 \\
\hline
$u$     & $>26.5 (3\sigma)$         &               &           &               &               &      \\        
$g$     & $25.9\pm0.3$    &         &           &               &               &      \\
K$_\mathrm{s}$ & $>23.3 (3\sigma)$  &               &           &               &               &      \\
\hline
\hline
\end{tabular}
\end{center}
\end{table*}

%\begin{table}
%\caption{Parameters from the Galfit fits\label{Galfit:results}}
%\begin{center}
%\begin{tabular}{llllll}
%\hline
%\hline
%Band  & mag &  r$_{eff}$ & PA & b/a & n \\
%      &     & (arcsec) & (deg) &   &  \\
%F606w & 25.52$\pm$0.13 & 0.149$\pm$0.024 & -37$\pm$6 & 0.36$\pm$0.09 & 0.78$\pm$0.37 \\
%F105w & 24.63$\pm$0.09 & 0.160$\pm$0.035 & -44$\pm$9 & 0.28$\pm$0.19 & 0.14$\pm$0.66 \\
%F160w & 23.69$\pm$0.06 & 0.174$\pm$0.016 & -47$\pm$6 & 0.43$\pm$0.09 & 1.10$\pm$0.41 \\
%\hline
%\hline
%\end{tabular}
%\end{center}
%\end{table}

% extinction correction: NED Schlegel
% u 0.104 g0.081 z 0.031 J 0.017 H 0.011 --> F105W 0.024
%ACS F606W  (0.59)  0.065
%WFC3 F606W 0.061
%WFC3 F105W  (1.04)  0.024
%WFC3 F160W  (1.53)  0.013

The $z=2.583$ DLA galaxy has a disk-like morphology with a Sersic index
consistent with 1. The galaxy is compact with a circularized radius of only
$0.11\arcsec$ corresponding to $0.9$ kpc.

\subsection{NOT/Alfosc imaging}
\label{out_of_alf}

We use $mag_{auto}$ in SExtractor \citep{1996A&AS..117..393B} to measure the total fluxes of SDSS stars in the u- and g-band images of the field of Q\,0918$+$1636, which we use to derive the zeropoints.  
In order to do photometry of the DLA galaxy counterpart we first did PSF
subtraction using the same procedure as in similar previous studies
\citep[e.g.,][]{1993A&A...270...43M,1999MNRAS.305..849F,2000A&A...358...88F}.
Magnitudes were measured in circular apertures. Again, Galactic extinction
corrections are taken from the \cite{2011ApJ...737..103S} maps. In the g-band
we measure an AB magnitude of $25.9\pm0.3$ and in the u-band we do not detect
the DLA galaxy down to a 3$\sigma$ detection limit of 26.5 (in a 2 arcsec
diameter aperture). %There is a faint blue source 1.0 arcsec North East of the
%QSO position, which appears to be real and not a residual from the 
%PSF-subtraction. If real this
%must be an unrelated interloper (we do not see emission
%lines from this source at the expected positions for any of the two DLAs in the
%X-Shooter spectra).  

\subsection{NOT/NOTCam imaging}
\label{out_of_nc}

For the NOTCam/K$_\mathrm{s}$-band we determined the zeropoint with stars from the 2MASS catalog \citep{2006AJ....131.1163S}. 
On the combined image in the K$_\mathrm{s}$-band we again subtracted the PSF of the QSO
using a PSF determined from stars in the field.  The residual image does not
contain significant emission at the position of the $z=2.583$ DLA galaxy. 
There is residual flux at the expected position at the $2\sigma$ significance level,
but we conservatively report a 3$\sigma$ detection limit of 23.3
(on the AB system) measured in a 2 arcsec diameter aperture.
%From the SED fit we expect the 
%Ks-band magnitude of the $z=2.58$ DLA to be around $Ks\sim21.5$ (Vega). 
%Using the NOTCam exposure time calculator we estimate that it is not possible
%to detect the galaxy within the obtained total integration time, even if there
%was no QSO to be subtracted.

\subsection{SED fitting}
\label{sed_fits}

We fit stellar population synthesis models to the six broad-band photometric
points from the {\it HST} and NOT imaging listed in Table~\ref{Galfit_results},
to derive the stellar mass, age and star formation rate with the same procedure
as in \citet{2013MNRAS.433.3091K}. In summary the fitting code uses the stellar
population templates from \citet{BC03} convolved with a large Monte Carlo
library of star formation histories (exponential plus random bursts) assuming a
\citet{Chabrier2003} IMF. Dust is added following the two-component model of
\citet{Charlot2000}, with the parameters being the total optical depth,
$\tau_{\mathrm{v}}$, and the fraction of dust\footnote{For details on the prior
distribution of the SFH and dust parameters see \citet{Salim2005}} contributed
by the ISM, $\mu$.  The metallicity is restricted to solar as inferred from the
absorption analysis, but we find consistent results when using the full range
of the models between 20\% and 2.5 times solar. We then adopt a Bayesian
approach by comparing the observed magnitudes to the ones predicted by all the
models in the library, and we construct the probability density functions of
stellar mass, mean luminosity-weighted stellar age, and star formation rate.
Fortunately, none of the filters contains any of the strong emission line and
hence we do not include emission lines in the fits. The results of the SED-fits
are provided in Table~\ref{SED_results}.

\begin{table}
\caption{Results from the SED fitting\label{SED_results}}
\begin{center}
\begin{tabular}{lllllll}
\hline
\hline
Parameter & Value \\
\hline
%textbf\{Best-fit model} & textbf\{???} \\
Age~[Myr] & $233^{+268}_{-125}$ \\ 
$E_{B-V}$~[mag] & $0.38^{+0.16}_{-0.12}$ \\
$A_{V}$~[mag] & $1.54^{+0.72}_{-0.56}$ \\
$M_{\star}$~[$10^{9}~M_{\odot}$] & $12.6^{+6.1}_{-2.9}$ \\ 
SFR~[$M_{\odot}yr^{-1}$]$^1$ & $27^{+20}_{-9}$ \\
\hline
\end{tabular}
\end{center}
$^1$ Averaged over 1 Gyr, but averaged over a shorter timescale of 10 Myr we get a similar value within the errors. 
\end{table} 

\begin{figure}
\includegraphics[width=0.48\textwidth]{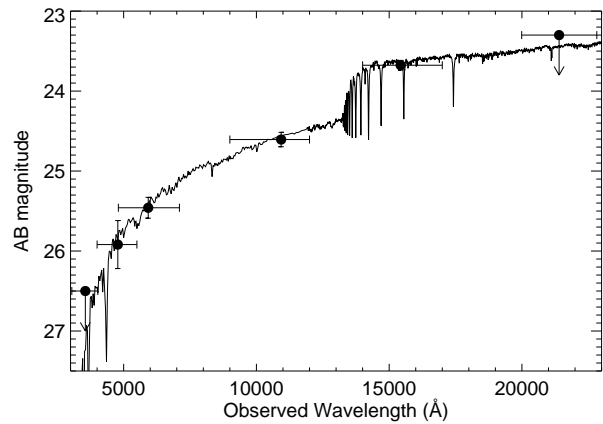}
\caption{The broad-band SED of the $z=2.583$ DLA galaxy, 
comprising of the Alfosc
u- and g-bands, HST/WFC3 F606W-, F105W- and F160W-bands and the NOTCam K$_\mathrm{s}$-band, is shown as black
points. The best-fit model is shown with a full-drawn line.
\label{LePhare_figure}}
\end{figure}
\subsection{VLT/X-Shooter spectroscopy}
\label{2583shoot}
%\label{shoot_out}

%Due to the proximity of the DLA galaxies to the background QSO, their spectral
%point spread functions (sPSFs) are blended. In order to extract the spectrum of
%the foreground galaxies, we subtract the sPSF of the QSO, with the method which
%was used in \citet{2010MNRAS.408.2128F}. This is only possible for the emission
%lines from the foreground galaxies as any continuum emission will be subtracted
%together with the continuum from the background QSO.

\begin{figure}
\includegraphics[width=0.48\textwidth]{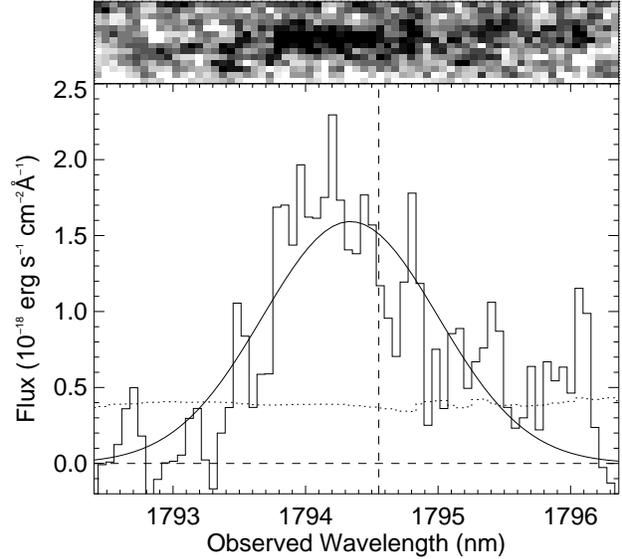}
\caption{The [\ion{O}{iii}]\,$\lambda$5007 emission line of the 
$z=2.583$ DLA galaxy. The top panel shows the 
2D spectrum and the bottom panel shows the 1D spectrum.
The vertical dashed line indicates the predicted 
position of the line based on the absorption line redshift
of the DLA. The observed line has a centroid that is blue shifted
by 36 km s$^{-1}$ relative to the absorption redshift. The FWHM 
of the line based on a Gaussian fit is $256\pm23$ km s$^{-1}$
(uncorrected for the spectroscopic resolution of 45 km s$^{-1}$).
\label{fig:oiii2.58}}
\end{figure}

The galaxy which is responsible for the $z=2.583$ DLA is located at a projected
distance of $1\farcs98$ from the QSO, which corresponds to 10 pixels on the
spatial axis of the 2D spectrum. Given the good seeing of $0\farcs8$ the two
objects are well separated. Thus, we do not need to subtract the continuum of
the QSO. At the spatial position where the DLA galaxy is located, there is very
little continuum flux, but the [\ion{O}{iii}]\,$\lambda$5007 emission line of
the $z=2.583$ DLA galaxy is clearly visible. The [\ion{O}{iii}]\,$\lambda$5007
line is typically the line detected at highest S/N ratio for galaxies at 
similar redshifts
\citep[e.g.,][]{2011MNRAS.413.2481F}.  We then extract a 1D spectrum of the
$z=2.583$ DLA galaxy similarly as for the QSO spectrum, but this time
using a gaussian SPSF with a FWHM of $0\farcs8$, as there is not enough
continuum signal to determine the SPSF. The [\ion{O}{iii}]\,$\lambda$5007
emission line is clearly located in wavelength regions without sky-line
residuals. For this line the flux is determined by summing up the flux in the
1D spectrum. In Fig~\ref{fig:oiii2.58} we show the
[\ion{O}{iii}]\,$\lambda$5007 line in 1 and 2 dimensions. The redshift
determined from the [\ion{O}{iii}]\,$\lambda$5007 line is
$z=2.58277\pm0.00010$, which is $36\pm20$ km s$^{-1}$ (where the uncertainty
also includes the uncertainty on the absorption redshift) blueshifted compared
to the centre of the low-ionisation absorption lines
\citep{2011MNRAS.413.2481F}.

\begin{figure}
\includegraphics[width=0.48\textwidth]{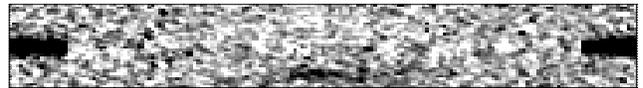}
\caption{A wider region around the [\ion{O}{iii}]\,$\lambda$5007 
line of the $z=2.583$ DLA galaxy after
subtraction of the QSO SPSF (the boundaries of where the QSO continuum has 
been subtracted can be seen at each end of the figure).
As seen here, there is no evidence for
[\ion{O}{iii}]\,$\lambda$5007 emission at small impact parameter. A 
[\ion{O}{iii}]\,$\lambda$5007 line 3 times fainter than the detected line
would have been detected even if superposed on the QSO
trace.
\label{fig:oiii2.58_qsosub}}
\end{figure}

To search for emission at lower impact parameter we 
subtract the QSO continuum
following the procedure described in \citet{2010MNRAS.408.2128F}.
In Fig.~\ref{fig:oiii2.58_qsosub} we show a wider region around the 
[\ion{O}{iii}]\,$\lambda$5007 line from the $z=2.583$ DLA galaxy after 
subtraction of the SPSF of the QSO. There is
no evidence for emission at smaller impact parameters.

The H$\alpha$, H$\beta$, [\ion{O}{iii}]\,$\lambda$4960, and 
[\ion{O}{ii}]\,$\lambda$3727 lines are visible but detected at lower 
S/N ratio. We derive fluxes for these lines by fixing the redshift and
width from the [\ion{O}{iii}]\,$\lambda$5007 line. The resulting line fluxes are
provided in Table~\ref{FluxTable}. In particular H$\alpha$ is very uncertain
as it is located far in the red end of the K-band where the sky-background is 
very high. 
We use the emission-line ratio R$_{23}$ (originally defined by
\cite{Pagel1979}) to derive the oxygen abundance for the system. The index is
defined as the ratio of [\ion{O}{ii}]\,$\lambda$3727 and [\ion{O}{iii}]\,$\lambda\lambda$
4959,5007 to H$\beta$. The R$_{23}$ metallicity indicator 
is double-valued. Moreover, the
calibration of the line ratio depends on the ionization parameter, which also
depends on metallicity. We therefore solve the problem iteratively by use of
the line ratio O$_{32}$ as an indicator of the ionization parameter. By using
the calibration of \cite{Kobulnicky2004} to infer the metallicity, we obtain
the following two values: the upper branch solution is $12 + \log(\rm{O/H}) =
8.8\pm0.2$, and the lower branch solution is $12 + \log(\rm{O/H}) = 8.2\pm0.2$.
We consider the upper branch solution most likely in this case given the
other properties of the system (absorption metallicity, luminosity, mass),
but we cannot establish this on the basis of the emission lines alone.

For the H$\alpha$ emission line the observed line flux corresponds to a
luminosity of $L_{\mathrm{H}\alpha} = 1.5\pm0.5\times10^{42}$~erg~s$^{-1}$.
Converting the luminosity into SFR using \citet{Kennicutt98} gives
SFR$_{\mathrm{H}\alpha} = 13\pm5~\mathrm{M}_{\odot}$~yr$^{-1}$.  Converting to
the assumed Chabrier IMF \citep{2007ApJS..173..256T} we find
SFR$_{\mathrm{H}\alpha} = 8\pm3~\mathrm{M}_{\odot}$~yr$^{-1}$.  Correcting for
the extinction inferred from the SED fitting this corresponds to
$22\pm7~\mathrm{M}_{\odot}$~yr$^{-1}$ for the Chabrier IMF, consistent with the
SFR derived from the SED fitting in Table~\ref{SED_results}.

Due to the increased S/N ratio of the spectrum of Q0918+1636 we are also
able to detect much weaker absorption lines than in
\citet{2011MNRAS.413.2481F}.  An example is \ion{Ti}{ii}\,$\lambda$1910 for which we
measure an observed equivalent width of 0.102$\pm$0.014 \AA \ corresponding to
a metallicity of -0.98$\pm$0.05 implying that Titanium is depleted by close to
1 dex. This is consistent with observations of Titanium in the local group
where Titanium is found to be highly depleted onto dust grains
\citep[e.g.,][]{2010MNRAS.404.1321W}. It is also consistent with the large
depletion of Fe, Mn and Cr \citep{2011MNRAS.413.2481F}.

\section{The $\lowercase{z}=2.412$ DLA Galaxy}
\label{z2412}

%In this section we present our analysis of the
%$z=2.412$ DLA and its galaxy counterpart.

\subsection{Absorption line analysis}

\begin{figure}
\begin{center}
	\includegraphics[width=0.48\textwidth]{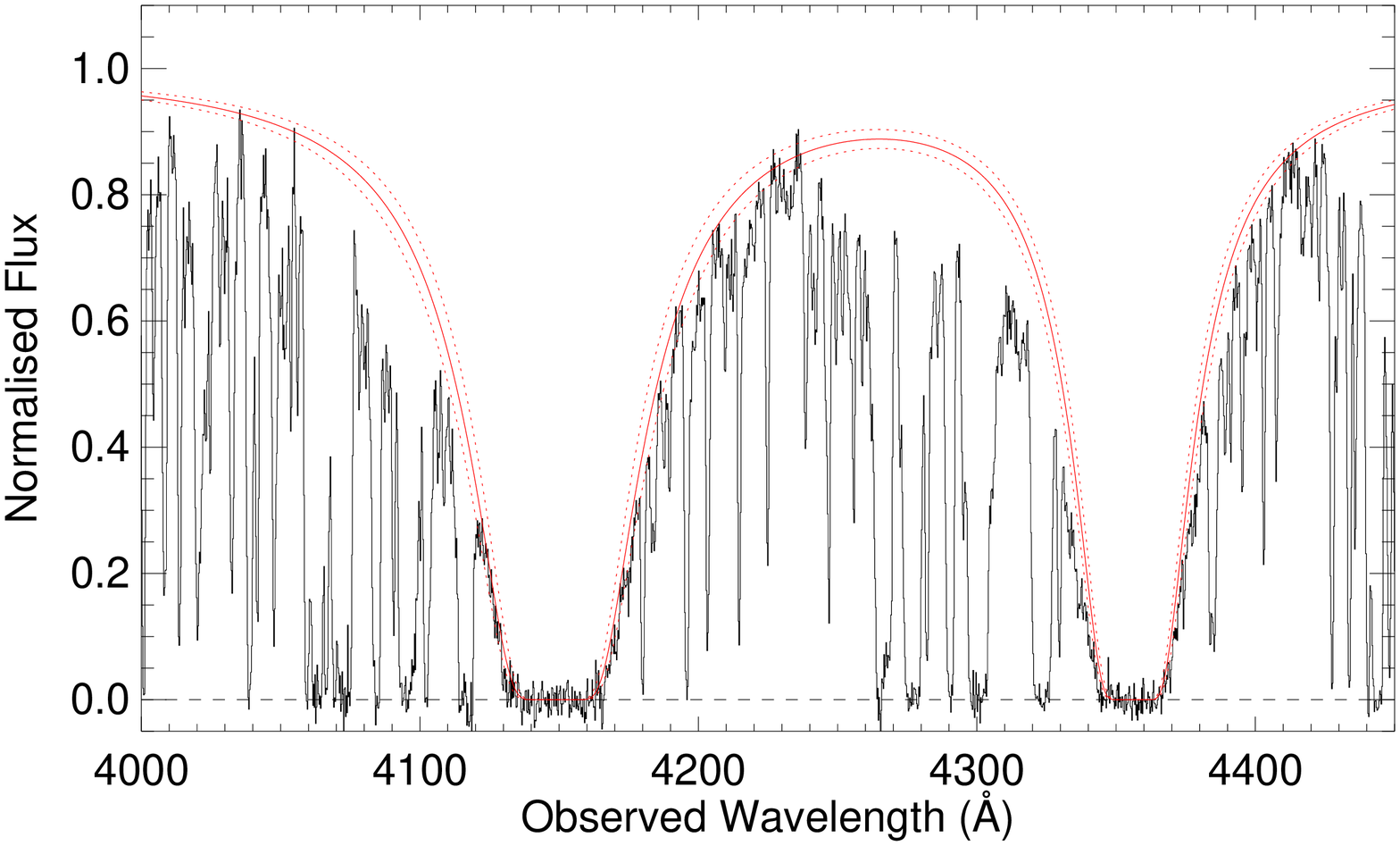}
\end{center}
\caption{Voigt profile fits to the two
DLA lines. For the $z=2.583$ DLA at $\lambda=4355$ \AA \ the fit is 
from \citet{2011MNRAS.413.2481F}. 
The derived column density for the $z=2.412$ DLA is $\log{N/\mathrm{cm^{-2}}} = 21.26\pm 0.06$.
}
\label{figure:dla}
\end{figure}

The original reason for targeting this QSO was the presence of a metal-strong
DLA at $z=2.412$. 
To characterize the absorption line properties of the $z=2.412$ DLA we performed
Voigt-profile fitting of the \ion{H}{i} and metal absorption lines.
For the DLA we derive an \ion{H}{i} column density of $\log{N/\mathrm{cm^{-2}}} = 21.26\pm 0.06$
(Fig.~\ref{figure:dla}).

\begin{figure}
\begin{center}
	\includegraphics[width=0.45\textwidth]{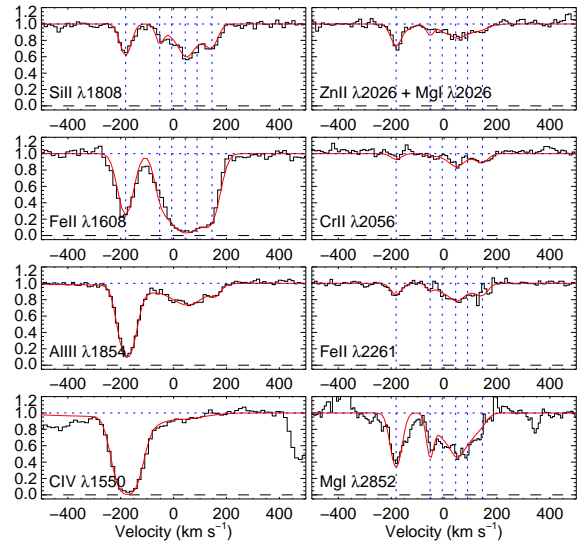}
\end{center}
\caption{Voigt-profile fits to metal lines from the $z=2.412$ DLA. The 
zero-point for the velocity scale is defined from the centroid of the 
[OIII]\,$\lambda$5007 emission line. Data are shown in black and the model-fits in red. In the left column we list a weak and a
strong low-ionisation line (\ion{Si}{ii} and \ion{Fe}{ii}) and a medium- and
high-ionisation line (\ion{Al}{iii} and \ion{C}{iv}) to illustrate the very
different ionization states of the different sub-components. The vertical dotted
lines mark the velocities for the sub-components in the fit to the
low-ionisation lines. In the right column we show the low-ionisation lines
from \ion{Zn}{ii}, \ion{Cr}{ii}, \ion{Fe}{ii} and \ion{Mg}{i}.
The zeropoint for the velocity scale is defined by the centroid (first
moment) of the \ion{Si}{ii}\,1808 line.}
\label{figure:metals}
\end{figure}

Voigt-profiles were fitted to several metal lines using the VPfit
software\footnote{http://www.ast.cam.ac.uk/~rfc/vpfit.html}, assuming
turbulence-dominated internal motion in the system. The redshift measured from
the low-ionization (\ion{Si}{ii}, \ion{Fe}{ii}, \ion{Zn}{ii}, \ion{Cr}{ii},
\ion{Mn}{ii}) absorption lines is $z=2.4121\pm0.0002$.  Redshifts and velocity
dispersions were tied for each of the individual components of the
low-ionisation lines. High-ionisation lines were fitted independently.  The
low-ionization absorption was found to be best fitted by six distinct
components.  A plot of the fit results is shown in Fig.~\ref{figure:metals},
and inferred column densities are shown in Table \ref{table:metals}. Integrated
metallicities [M/H] based on low-ionization absorption are $-0.6\pm0.2$,
$-0.6\pm0.2$, $-1.2\pm0.2$, $-1.2\pm0.2$ and $-1.3\pm0.2$ for
\ion{Zn}{ii}, \ion{Si}{ii}, \ion{Cr}{ii}, \ion{Fe}{ii} and \ion{Mn}{ii},
respectively. In Fig.~\ref{figure:metals} we also show the intermediate 
and high-ionisation  lines from \ion{Al}{iii} and \ion{C}{iv}. These lines are fitted
independently and in this case only 5 sub-components suffice. It is striking that the $v<0$ km s$^{-1}$
absorption is strongest for \ion{Al}{iii} and \ion{C}{iv} whereas the strongest 
low-ionization absorption is at $v>0$ km s$^{-1}$.

The resolution of X-Shooter is, as dicussed in several earlier
works, not ideal for robust Voigt-profile fitting
\citep[e.g.,][]{2010MNRAS.408.2128F,Noterdaeme12,2013A&A...557A..18K},
but for our purposes of establishing that the system is metal-rich and for
inferring the velocity width of the absorption, the data are sufficient. 

We also determine the velocity width of the low-ionisation absorption following
the prescription of \citet{2006A&A...457...71L}. Here we find $\Delta v = 349$
km s$^{-1}$ and 352 km s$^{-1}$ for \ion{Fe}{ii},$\lambda$2260 and 
\ion{Si}{ii},$\lambda$1808,
respectively.

In conclusion, the metallicity of the system is well above our
target selection criterion of 0.1 Z$_{\sun}$ and there is evidence for
substantial depletion of refractory elements on dust grains.

\begin{table}
\caption {Ionic column densities in the 6 individual line components of the DLA system at
$z_{\rm abs}=2.412$.\label{table:metals}}
\begin{center}
\begin{tabular}{llll}
\hline
\hline
Ion & Transition & $\log N\pm\sigma _{\log N}$ & $b\pm\sigma _b$\\
    & lines used &                            & (km s$^{-1}$)  \\
\hline
\multicolumn{4}{l}{$z_{\rm abs}=2.41033$}  \\
MgI  & 2026,2852        & 12.60$\pm$0.05  &  38$\pm$1  \\
SiII & 1808             & 15.40$\pm$0.05  &  38$\pm$1  \\
MnII & 2576,2594,2606   & 12.79$\pm$0.05  &  38$\pm$1  \\
FeII & 1608,1611,2249,2260   & 14.70$\pm$0.05 & 38$\pm$1  \\
ZnII & 2026, 2062       & 12.77$\pm$0.05  &  38$\pm$1  \\
CrII & 2056,2062,2066   & 12.80$\pm$0.10  &  38$\pm$1  \\
\hline
\multicolumn{4}{l}{$z_{\rm abs}=2.41179$}  \\
MgI  & 2026,2852        & 12.34$\pm$0.07  &  20$\pm$3  \\
SiII & 1808             & 15.0$\pm$0.1  &  20$\pm$3  \\
MnII & 2576,2594,2606   & 11.9$\pm$0.2  &  20$\pm$3  \\
FeII & 1608,1611,2249,2260 & 14.29$\pm$0.08 & 20$\pm$3  \\
ZnII & 2026, 2062      & 12.3$\pm$0.1  &  20$\pm$3  \\
CrII & 2056,2062,2066  & 12.3$\pm$0.4  &  20$\pm$3  \\
\hline
\multicolumn{4}{l}{$z_{\rm abs}=2.41231$}  \\
MgI  & 2026,2852        & 12.1$\pm$0.1  &  22$\pm$6  \\
SiII & 1808             & 15.3$\pm$0.7  &  22$\pm$6  \\
MnII & 2576,2594,2606   & 12.6$\pm$0.7  &  22$\pm$6  \\
FeII & 1608,1611,2249,2260 & 14.8$\pm$0.7 & 22$\pm$6  \\
ZnII & 2026, 2062      & 12.4$\pm$0.7  &  22$\pm$6  \\
CrII & 2056,2062,2066  & 12.9$\pm$0.8  &  22$\pm$6  \\
\hline
\multicolumn{4}{l}{$z_{\rm abs}=2.41289$}  \\
MgI  & 2026,2852        & 12.4$\pm$0.5  &  37$\pm$11  \\
SiII & 1808             & 15.4$\pm$0.5  &  37$\pm$11  \\
MnII & 2576,2594,2606   & 12.6$\pm$0.6  &  37$\pm$11  \\
FeII & 1608,1611,2249,2260 & 14.9$\pm$0.5 & 37$\pm$11  \\
ZnII & 2026,2062        & 12.5$\pm$0.5  &  37$\pm$11  \\
CrII & 2056             & 13.2$\pm$0.3  &  37$\pm$11  \\
\hline
\multicolumn{4}{l}{$z_{\rm abs}=2.4134$}  \\
MgI  & 2026,2852        & 12.4$\pm$0.6  &  45$\pm$28  \\
SiII & 1808             & 15.3$\pm$0.7  &  45$\pm$28  \\
MnII & 2576,2594,2606   & 12.6$\pm$0.7  &  45$\pm$28  \\
FeII & 1608,1611,2249,2260 & 14.8$\pm$0.7 & 45$\pm$28  \\
ZnII & 2026, 2062      & 12.4$\pm$0.7  &  45$\pm$28  \\
CrII & 2056,2062,2066  & 12.9$\pm$0.8  &  45$\pm$28  \\
\hline
\multicolumn{4}{l}{$z_{\rm abs}=2.41404$}  \\
MgI  & 2026,2852        & 11.5$\pm$1.0  &  28$\pm$3  \\
SiII & 1808             & 15.2$\pm$0.2  &  28$\pm$3  \\
MnII & 2576,2594,2606   & 12.5$\pm$0.2  &  28$\pm$3  \\
FeII & 1608,1611,2249,2260 & 14.7$\pm$0.2 & 28$\pm$3  \\
ZnII & 2026, 2062      & 11.9$\pm$0.4  &  28$\pm$3  \\
CrII & 2056,2062,2066  & 13.0$\pm$0.1  &  28$\pm$3  \\
\hline
\end{tabular}
\end{center}
\end{table}

Further details on the analysis of this system can be found in 
\citet{ThorsenThesis:2011}.

\subsection{The galaxy counterpart}
In \citet{2011MNRAS.413.2481F} no emission was found from the galaxy
counterpart of this absorber. From the spectrum taken with the slit at the
position angle of 162$^\mathrm{o}$ East of North we find that the galaxy seen
at the bottom of Fig.~\ref{fig:slits} is at a lower redshift of $z=0.987$ based
on the detection of the [\ion{O}{ii}]\,$\lambda$3727 doublet and the
[\ion{O}{iii}]\,$\lambda$5007 line. In Fig.~\ref{fig:mosaic} there is no
obvious other source at smaller impact parameters except the counterpart of the
$z=2.583$ DLA. One possibility is that the source is a faint galaxy at a small
impact parameter.  In the lower left sub-panel of the upper panel in
Fig.~\ref{fig:mosaic} there is a hint of a source at a position angle of about
130$^\mathrm{o}$ East of North (marked with an arrow). Opposite to this is a
ring-like residual consistent with what one would expect if the centroid of the PSF
has been shifted slightly to the lower left by the presence of the foreground
galaxy. We have attempted to include such an additional source in the G{\small
ALFIT} modelling, but without success. We have done one test to gauge the
reality of this potential source. Using G{\small ALFIT} with rotated PSFs
(45$^\mathrm{o}$, 90$^\mathrm{o}$, 135$^\mathrm{o}$, etc.), we find that the
dimples surrounding the PSF core are either well subtracted or poorly
subtracted (or, in general, partially subtracted) depending on the angle of
rotation. Obviously, the main effect is that the diffraction spikes change such
that with a rotation one sees the original spikes as a positive residual and a
negative residual oriented along the diffraction spikes of the rotated PSF.
However, the residual to the south-east, labeled in the image, remains fixed
with roughly the same shape and brightness. In the end we are convinced that
the residual is due to a real source and not an artifact from poor PSF
subtraction. We note that the residual could also be related to the
host galaxy of the QSO. 
The tail extending below the QSO is a result of Charge
Transfer Inefficiency.

\begin{figure}
\includegraphics[width=0.48\textwidth]{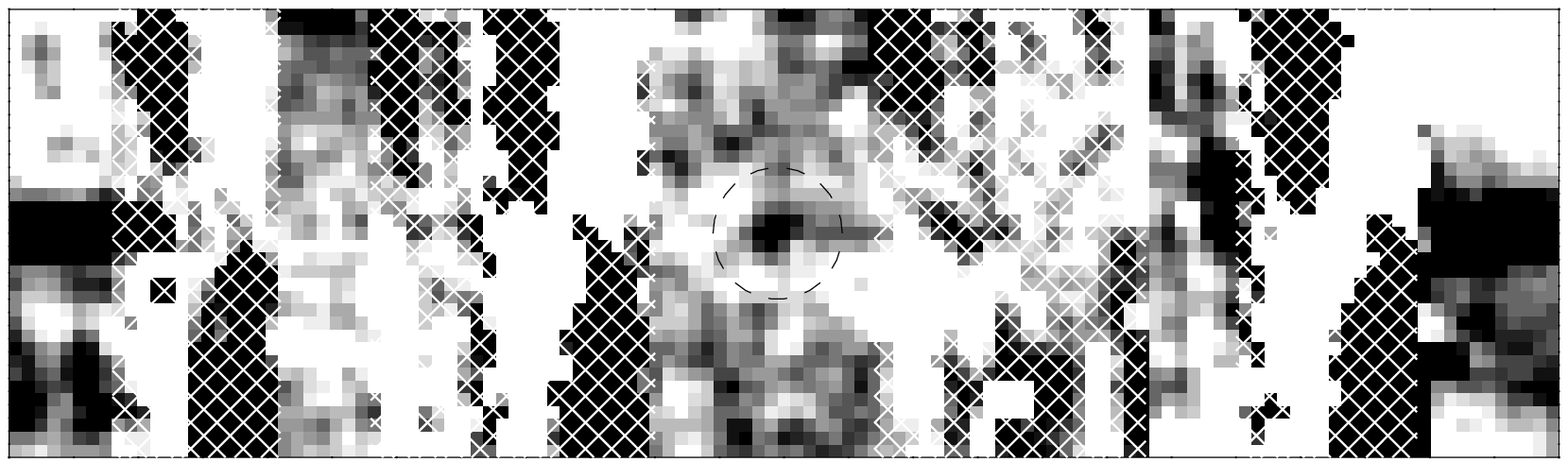}
\includegraphics[width=0.48\textwidth]{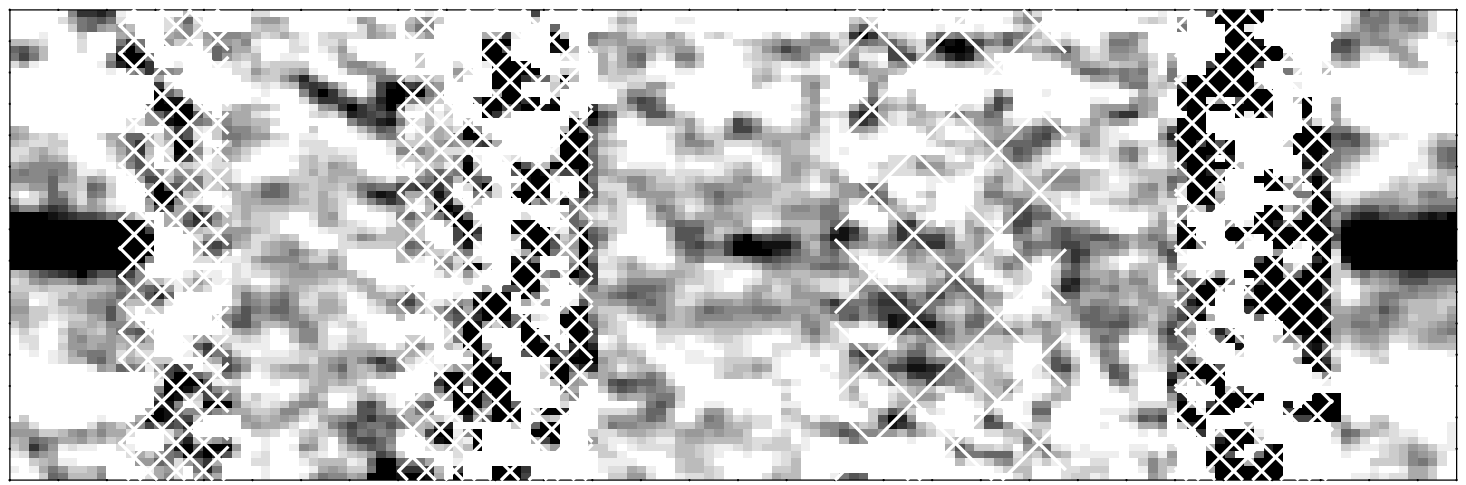}
\caption{The region around the position in the spectrum
where the [\ion{O}{iii}]\,$\lambda$5007 emission line from the $z=2.412$
DLA galaxy is expected to be observed. 
The upper panel is based on a stack of all the data taken up to
and including 2012 and the lower panel is based on the 3 hr 
observation from March 2013. The QSO
continuum can be seen in the left and right extremes of the figures, but
in between it has been subtracted. Regions with sky-lines from airglow
have been marked with a double-hatched pattern. The
[\ion{O}{iii}]\,$\lambda$5007 emission line from the $z=2.412$ DLA galaxy is
marked with a dashed circle in the upper panel.
\label{fig:OIII}}
\end{figure}

In the case of a very small impact parameter we expect potential emission lines from the galaxy to be included in all
the slits shown in Fig.~\ref{fig:slits}. We therefore co-added all the
2-dimensional spectra obtained up to 2012 and performed a SPSF-subtraction as described in
\citet{2010MNRAS.408.2128F}. In the upper panel of Fig.~\ref{fig:OIII} we show the region around
the position in the spectrum where the [\ion{O}{iii}]\,$\lambda$5007 emission line is
expected to fall. We tentatively, at about 3.5$\sigma$ significance, detect
an emission line at the expected position. 
The spectrum we obtained in March 2013
is substantially better due to better observing conditions and the use of the
nodding observing template. In the lower panel of Fig.~\ref{fig:OIII} we 
show the [\ion{O}{iii}]\,$\lambda$5007 emission line from the $z=2.412$ DLA 
now at a higher S/N ratio. The line is very narrow with a FWHM of 67$\pm$12
km s$^{-1}$, which is only slightly larger than the resolution of 45 km s$^{-1}$. 
Corrected for the resolution the FWHM is 50 km s$^{-1}$.
The redshift determined from the line is $z=2.4128\pm0.0002$, 
which is redshifted $38\pm25$ km s$^{-1}$ relative
to the mean absorption redshift, but of course well within the full $\sim$350 
km s$^{-1}$ velocity extent of the low-ionisation absorption.

\begin{table}
\caption{Measured emission line fluxes\label{FluxTable}}
\begin{center}
\begin{tabular}{l c r c c}
\hline
\hline
Transition & Wavelength$^{(1)}$ & Flux$^{(2)}$ & FWHM$^{(3)}$ \\
\hline
$z=2.412$ DLA & & & \\
$[$O{\sc \,iii}$]$  &   5006.84 &   4.1$\pm$1.1 & 50$\pm$12  \\
\hline
$z=2.583$ DLA & & & &\\
$[$O{\sc\,ii}$]$   &   3726.03,\,3728.82 &   25$\pm$4 & ~~ \\
$[$O{\sc\,iii}$]$  &   4958.92 &   11$\pm$3 & ~~ \\
$[$O{\sc\,iii}$]$  &   5006.84 &   25$\pm$3 & $252\pm23$~~ \\
H$\beta$   &   4861.325 &   13$\pm$3 & ~~ \\
H$\alpha$   &   6562.80 &   27$\pm$10 & ~~ \\

\hline
\end{tabular}
\end{center}
$^{(1)}$ Transition rest frame wavelength in $\mathrm{\AA}$.\\
$^{(2)}$ Flux in units of $10^{-18}$erg s$^{-1}$ cm$^{-2}$.\\
$^{(3)}$ Line width at FWHM in units of km~s$^{-1}$ corrected for the instrumental 
resolution of 45 km s$^{-1}$.
\end{table}

We do not detect other lines, but [\ion{O}{iii}]\,$\lambda$5007 is the line expected to be detected at highest
significance at these redshifts 
\citep[e.g.,][]{2010MNRAS.408.2128F,2012ApJ...758...46K,2013A&A...557A..18K}
and the non-detection of the other lines is expected on S/N grounds.

The impact parameter is consistent with 0 and a conservative upper limit is 
0.25 arcsec corresponding to 2.0 kpc.

\section{Discussion}
\label{Discussion_Section}

In this paper we have presented new observations of the two DLAs towards
Q\,0918+1636 and their galaxy counterparts. The galaxy counterpart of the $z=2.583$
DLA was discovered previously \citep{2011MNRAS.413.2481F}, whereas the
discovery of the counterpart of the $z=2.412$ DLA is first
reported here.

\subsection{The $z=2.583$ DLA Galaxy}

For the $z=2.583$ DLA Galaxy we have the largest amount of information: 
detection of severeal strong emission lines and a clear detection of the 
galaxy in the {\it HST} images.
We can use our information about the size of the galaxy and the kinematics, as
probed by the [\ion{O}{iii}]\,$\lambda$5007 emission line of the $z=2.583$ DLA galaxy to
get an estimate of the dynamical mass of the system. As in \citet{2013MNRAS.433.3091K} 
we follow the method
described in \citet{Rhoads2013} to estimate the dynamical mass given the
measured size and velocity dispersion:
$$ M_{\mathrm{dyn}} \approx \frac{4\,\sigma^2\, r_{\rm eff}}{G \sin^2(i)} ~,$$
\\
where $i$ denotes the inclination of the system with $i=90^{\mathrm{o}}$ being
edge on and $G$ is the gravitational constant. In order to estimate the velocity dispersion of the system we use the
FWHM of the emission lines as a probe of the integrated gas-kinematics of the
system. We then use the width of the [\ion{O}{iii}]\,$\lambda$5007 line to estimate the
velocity dispersion to be $\sigma=107\pm10$ km s$^{-1}$, and we adopt the size from the 
G{\small ALFIT} analysis:
$r_{\rm eff}=1.4$ kpc (see Table~\ref{Galfit_results}).

From the G{\small ALFIT} analysis we infer a (projected) axis ratio of the galaxy 
of $b/a=0.43$. The system may be described as disc-like, given the elongated shape, 
and the fact that we see a value of S\'ersic $n$ close to 1. We thus adopt a value of 
$\sin(i)=0.5$ and use the fitted half-light semi-major axis for our estimate of the 
dynamical mass of the system:
$M_{\mathrm{dyn}} \approx 6.0\pm1.3\times10^{10}$~$M_{\odot}$. This estimate should 
only be considered a rough approximation given the assumptions.

From our SED fit to the broad band imaging data, we obtain a stellar mass of
$M_{\star} \gtrsim10^{10}$$M_{\odot}$. We can use this measurement to test the
mass-metallicity relation for DLA systems
\citep{2006A&A...457...71L,2013MNRAS.430.2680M}. Using
the relation in \citet[][their eq. 6]{2013MNRAS.430.2680M} using as input solar 
metallicity
(i.e., assuming that $C_{[M/H]}=0$) we predict a stellar mass of 
$M_{\star} = 3 \times 10^{10}$~$M_{\odot}$. Given the substantial ($\sim0.38$ dex) scatter 
in their relation, the agreement between the prediction and our best fit 
stellar mass from the SED fit is good.

\subsection{The nature of DLA galaxies}

\begin{table*}
\caption{Comparison of the two DLA galaxies studied here with DLA galaxies at similar redshifts
from the literature. In the table we refer to the $z=2.412$ and $z=2.583$ DLA galaxies as 
DLA0918+1636-1 and DLA0918+1636-2, respectively.
 \label{Compare}}
\begin{center}
\begin{tabular}{lccrrrrrr}
\hline
\hline
DLA galaxy & $z_{\rm abs}$ & $\log{N/\mathrm{cm^{-2}}}$ & [M/H] & b$^1$ & FWHM([\ion{O}{iii}]) & $\Delta v_{90}$ & $\Delta v $(DLA-[\ion{O}{iii}]) &  ref \\
           &     &                           &       & (kpc) & (km s$^{-1}$)        & (km s$^{-1}$)                   & (km s$^{-1}$)   &     \\
\hline
DLA0918+1636-1  & 2.412 & 21.26 & $-0.6$ &   $<2$ &  50  & 350 & $-38\pm25$ & (1) \\ 
DLA0918+1636-2  & 2.583 & 20.96 &  $0.0$ & $16.2$ & 252  & 295 &  $-36\pm20$ & (1) \\ 
DLA1135-0010    & 2.207 & 22.10 & $-1.1$ &  $0.9$ & 120  & 186 & $9\pm10$  &  (2) \\
DLA2222-0946 & 2.353 & 20.65 & $-0.5$ &  $6.2$ & 115  & 185 & $25\pm20$ & (3) \\
N-14-1C    & 1.920 & 20.67 & $-0.4$ &  $8.3$  & 220  & 136 & $-150\pm20$ & (4) \\
N-14-2C    & 1.920 & 20.67 & $-0.4$ &  $10.6$ & 180  & 136 & $80\pm9$ &  (4) \\
DLA2243-60 & 3.330 & 20.65 & $-0.7$ &  $22.9$ & 320  & 173 & $200\pm20$ &  (5) \\
%N-16-1D    & 3.149 & 20.0 & $-0.56$,$-1.1$ & $19.0$& 55  & $-150\pm6$* & & (4) \\
\hline
\end{tabular}
\end{center}
\begin{flushleft}
(1) This work, \citet{2011MNRAS.413.2481F}; (2) \citet{Noterdaeme12}; (3) \citet{2010MNRAS.408.2128F,2013MNRAS.433.3091K}; 
(4) \citet{2005MNRAS.358..985W,2006A&A...457...71L};
(5) \citet{Bouche2012,2006A&A...457...71L,2013arXiv1306.0134B}, Noterdaeme, private communication.\\
$^1$ Recalculated from the values in the original references to be consistent with the assumed cosmology.
\end{flushleft}
\end{table*}

The two DLA systems studied here are, as other systems in our survey
\citep{2010MNRAS.408.2128F,2011MNRAS.413.2481F,2012MNRAS.424L...1K}, drawn from
the extreme high-metallicity end of the distribution and hence should not be
considered typical examples of DLA galaxies. In Table~\ref{Compare} we compare
the two systems and include also DLA galaxies from the literature for
comparison
\citep{2005MNRAS.358..985W,2010MNRAS.408.2128F,2011MNRAS.413.2481F,Bouche2012,Noterdaeme12,2013MNRAS.433.3091K}.
For consistency we re-calculate the velocity shifts for N-14-1C and
N-14-2C using the centroids of the low-ionisation lines corresponding to $z_{\rm abs}=1.9205$.
The galaxy counterpart of the $z=2.412$ DLA is the system with the
highest velocity extent of the low-ionisation absorption. However, the
FWHM of its [\ion{O}{iii}] emission is the lowest in the sample. This
indicates that additional influences than mass must be important in
determining the velocity width of the low-ionisation absorption. %at least for some sources.
One such possible influence is of course outflows. Another important reason for this may be the low impact
parameter, which implies that a larger fraction of the gravitational potential is probed by the 
line-of-sight. We also observe here, as in \citet{2012MNRAS.424L...1K}, that the systems with the highest \ion{H}{i} column densities have the smallest impact parameters.
It would be interesting to carry out detailed comparisons of the quantities in Table~\ref{Compare}
with simulations, e.g., similar to the works of \citet{2008MNRAS.390.1349P} and \citet{2013MNRAS.433.3103R}.

For the galaxy counterpart of the $z=2.583$ we can establish further
properties: It is a compact ($r_{\rm eff} = 1$ kpc), strongly star-forming
galaxy with a centroid 16.2 kpc away from the line-of-sight to the background
QSO. The galaxy photometry is well fitted by galaxy templates with ages up to
several 100 Myr. The ratio between the \ion{H}{i} gas scale length of this DLA
galaxy, as measured by its impact parameter, and the light scale length, as
measured by its half-light radii, is of order 10 as seen in previous cases of
DLA galaxy counterparts \citep{2002ApJ...574...51M,2013MNRAS.433.3091K}.  This
is very different from the situation in local galaxies, where the gas only
extends up to a few times the extension of the light
\citep{1981AJ.....86.1825B}.  Our data are deep, but the $(1+z)^4$ dimming of
surface brightness with redshift is a very strong effect. Hence, an important
question is whether the measured compact morphology is only due to central high
surface brightness regions embedded in lower surface brightness regions with
extension more similar to the \ion{H}{i} gas, but below the surface brightness
detection limit of our data.  Such \ion{H}{i} central high surface brightness
regions are also seen in local spiral galaxies \citep[e.g.,][]{Carollo97}.  The
issue of morphology of star-forming galaxies at these redshifts as inferred
from WFC3/IR data has been studied intensively by \citet{Law12a,Law12b} who
find that these systems are not rotationally supported disk galaxies.  Rather,
they appear to be predominantly unstable, dispersion-dominated, systems fueled
by rapid gas accretion which presumably later form extended rotationally
supported disks. They also argue that all these galaxies drive strong outflows
with more massive galaxies driving less highly ionized outflows. Compared to
their sample the $z=2.583$ DLA galaxy is in the upper third of the mass
distribution. For the $z=2.583$ galaxy the distances and ages are also consistent with
a wind scenario: for an age of 233 Myr a mean speed of $\sim70$ km s$^{-1}$ is
required to reach 16 kpc. Such winds speeds are well below what is seen in
nearby (more modest) winds \citep[e.g.,][]{2013MNRAS.430.3235M}.

\citet{2013arXiv1306.0134B} argue for a similar system of a galaxy counterpart
to a DLA at an even larger impact parameter (DLA2243-60 in Table~\ref{Compare})
that the gas causing the DLA absorption is in a cold inflow. In their case the
DLA metallicity is $-0.72$, which is too high to be pristine gas. Hence, also in this case, an outflow must have been important for
determining the properties of the system. 

\citet{Rafelski11} use statistical arguments to show that most DLAs must probe 
atomic gas with very low star-formation efficiencies. This would be consistent
with a picture where metals in this gas originates from a wind rather than 
having been formed in situ. 

The large impact parameter of the $z=2.583$ galaxy could also be related
to other processes like tidal stripping similar to what is seen in the
Magellanic stream \citep{2009ApJ...695.1382M}. As DLAs are \ion{H}{i}
cross-section selected such systems will have a higher probability of being
selected \citep[see also][]{2011MNRAS.418.1115R,2013arXiv1305.5849R}.
However, we note that the correlation between metallicity and impact parameter found by
\citet{2012MNRAS.424L...1K} would not obviously result from such a scenario and
we do not see evidence for a nearby galaxy that could have caused tidal stripping.

We note that none of the two galaxies have Ly$\alpha$ in emission. This may
help explain the many non-detections resulting from searches for DLA galaxies
in the previous few decades \citep[][and references
therein]{1995ApJ...451..484L,2004A&A...422L..33M,2010MNRAS.408.2128F}.

A coherent picture of DLAs and their relation to emission selected galaxies
could be the following: DLAs originate from the outskirts of galaxies with
properties (i.e., sizes, luminosities, stellar masses, metallicities)
within the range of star-forming Lyman-break galaxies at similar
redshift, but due to their cross-section selection they are more likely to be drawn
from the fainter end of the luminosity function than emission selected galaxies
\citep{1999MNRAS.305..849F,2002ApJ...574...51M,2008ApJ...683..321F,
2008ApJ...681..856R,2011MNRAS.412L..55R}. There is evidence that DLA galaxies
fulfill a metallicity-luminosity relation
\citep{2004A&A...422L..33M,2006A&A...457...71L,2008ApJ...683..321F,
2013MNRAS.430.2680M} and therefore high-metallicity DLAs are expected to have
galaxy counterparts more similar to typical emission-selected galaxies ( 
i.e., Lyman-break galaxies seen in ground-based surveys) than
DLAs in general which probably have extreme galaxy counterparts 
\citep{1999MNRAS.305..849F,2000ApJ...534..594H,2008ApJ...681..856R}. 
The galaxy counterparts of the two DLAs towards Q0918+1636
are consistent with this picture.

\subsection{Outlook}
Thanks to new sensitive near-IR spectrographs the study of galaxy counterparts
of $z>2$ DLAs has now opened \citep{2005MNRAS.358..985W,2010MNRAS.408.2128F,
2011MNRAS.413.2481F,Bouche2012,Noterdaeme12,2013MNRAS.433.3091K}. The
identification of intervening DLAs towards transient sources like Gamma-Ray Burst
afterglows have also led to the detection of a galaxy counterpart and this
approach hence also appears promising for the future \citep{2012A&A...546A..20S}.
At the
moment observations like these are limited to the bright counterparts of the
highest metallicity DLAs. With the advent of extremely large telescopes
equipped with advanced adaptive optics in the next decade, however, such
studies can be extended to the galaxy counterparts of more typical DLAs and
hence a more complete unification of absorption and emission studies of
high-$z$ galaxies is within reach.

\section*{Acknowledgments}

We thank the anomymous referee for a very help report and Steve Schulze for
comments on an earlier version of the manuscript.
The Dark Cosmology Centre is funded by the DNRF. JPUF acknowledges support
from the ERC-StG grant EGGS-278202. LC acknowledges the support of the EU under
a Marie Curie Intra-European Fellowship, contract GA-2010-247117.  AG
acknowledges support from the EU FP7/2007-2013 under grant agreement n. 267251
AstroFIt. The research of JRM is supported through a Royal Society University
Research Fellowship.  TK acknowledges support by the European Commission under
the Marie Curie Intra-European Fellowship Programme in FP7.

\def\aj{AJ}
\def\araa{ARA\&A}
\def\apj{ApJ}
\def\apjl{ApJL}
\def\apjs{ApJS}
\def\apss{Ap\&SS}
\def\aap{A\&A}
\def\aapr{A\&A~Rev.}
\def\aaps{A\&AS}
\def\mnras{MNRAS}
\def\nat{Nature}
\def\pasp{PASP}
\def\aplett{Astrophys.~Lett.}

\bibliographystyle{mn}

\end{document}